\documentclass[a4paper,11pt]{article}
\pdfoutput=1 

\usepackage{jinstpub} 

\usepackage[utf8x]{inputenc}
\usepackage{upgreek,xspace}
\def\Sigma{\ensuremath\upsigma\xspace}
\usepackage{siunitx}
\usepackage{mhchem}

\title{\boldmath A monolithic ASIC demonstrator for the Thin Time-of-Flight PET scanner}

\author[a]{P. Valerio,}
\author[b]{R. Cardarelli,}
\author[a]{G. Iacobucci,}
\author[a]{L. Paolozzi,}
\author[a]{E. Ripiccini,}
\author[a]{D. Hayakawa}
\author[b]{S. Bruno,}
\author[b]{A. Caltabiano,}
\author[c]{M. Kaynak,}
\author[c]{H. R\"{u}cker,}
\author[a,d]{M. Nessi}

\affiliation[a]{DPNC, Département de physique des particules et corpusculaire, \\Geneva}
\affiliation[b]{INFN, Sezione di Roma Tor Vergata,\\Roma}
\affiliation[c]{IHP, Leibniz-Institut f\"{u}r innovative Mikroelektronik,\\Frankfurt (Oder)}
\affiliation[d]{CERN,\\Geneva}
\emailAdd{pierpaolo.valerio@unige.ch}

\abstract{Time-of-flight measurement is an important advancement in PET scanners to improve image reconstruction with a lower delivered radiation dose.  This article describes the monolithic ASIC for the TT-PET project, a novel idea for a high-precision PET scanner for small animals.
The chip uses a SiGe Bi-CMOS process for timing measurements, integrating a fully-depleted pixel matrix with a low-power BJT-based front-end per channel, integrated on the same \SI{100}{\micro\meter} thick die. The target timing resolution is \SI{30}{\pico\second} RMS for electrons from the conversion of \SI{511}{\kilo\electronvolt} photons. A novel synchronization scheme using a patent-pending TDC is used to allow the synchronization of 1.6 million channels across almost 2000 different chips at picosecond-level.
A full-featured demonstrator chip with a 3$\times$10 matrix of 500$\times$\SI{500}{\micro\meter\squared} pixels was produced to validate each block. Its design and experimental results are presented here.​}

\keywords{Analogue electronic circuits, Digital electronic circuits, Front-end electronics for detector readout, Timing detectors, Pixelated detectors and associated VLSI electronics}

\begin{document}
\maketitle
\flushbottom

\section{The TT-PET project}
\label{sec:intro}
Conventional PET imaging techniques use scintillating crystals to detect two back-to-back photons produced by a positron-electron annihilation to determine where the annihilation occurred. Without additional information, the event is placed anywhere on the line of response between the two acquired signals and, with enough statistics, an accurate image can be reconstructed. The addition of a Time of Flight (TOF) measurement can restrict the initial placement of the interaction point on the line of response, reducing it to a segment. A more precise timing information corresponds to a shorter segment, resulting in a less noisy image, or in a reduced dose to the patient due to the smaller statistics required. In order to extract valuable information on the position of the annihilation point, a high TOF precision is required (at least \SI{200}{\pico\second}), as the particles travel at the speed of light.\\
Goal of the TT-PET (Thin TOF-PET) project is to build a novel small-animal PET scanner with a target of \SI{30}{\pico\second} RMS time resolution for photon detection\cite{proceedings}. This value is well beyond the state-of-the-art for time-of-flight PET systems\cite{stateoftheart}, and is obtained by a radically different approach compared to traditional scanners. Multiple layers of monolithic silicon pixel detectors and high-Z photon-converters are stacked to convert incoming photons and digitize hits, providing their 3D position and timing. Data are reconstructed off-line to correct for systematic offsets, discriminate coincidences and reconstruct the acquired image.\\
The TT-PET project is funded by the Swiss National Science Foundation. The front-end design was carried out by the University of Geneva and the INFN Rome Tor Vergata.

\section{System design aspects}
\label{sec:system}
\begin{figure}[htbp]
	\centering
	\begin{minipage}{0.50\textwidth}
		\centering
		\includegraphics[width=0.7\textwidth]{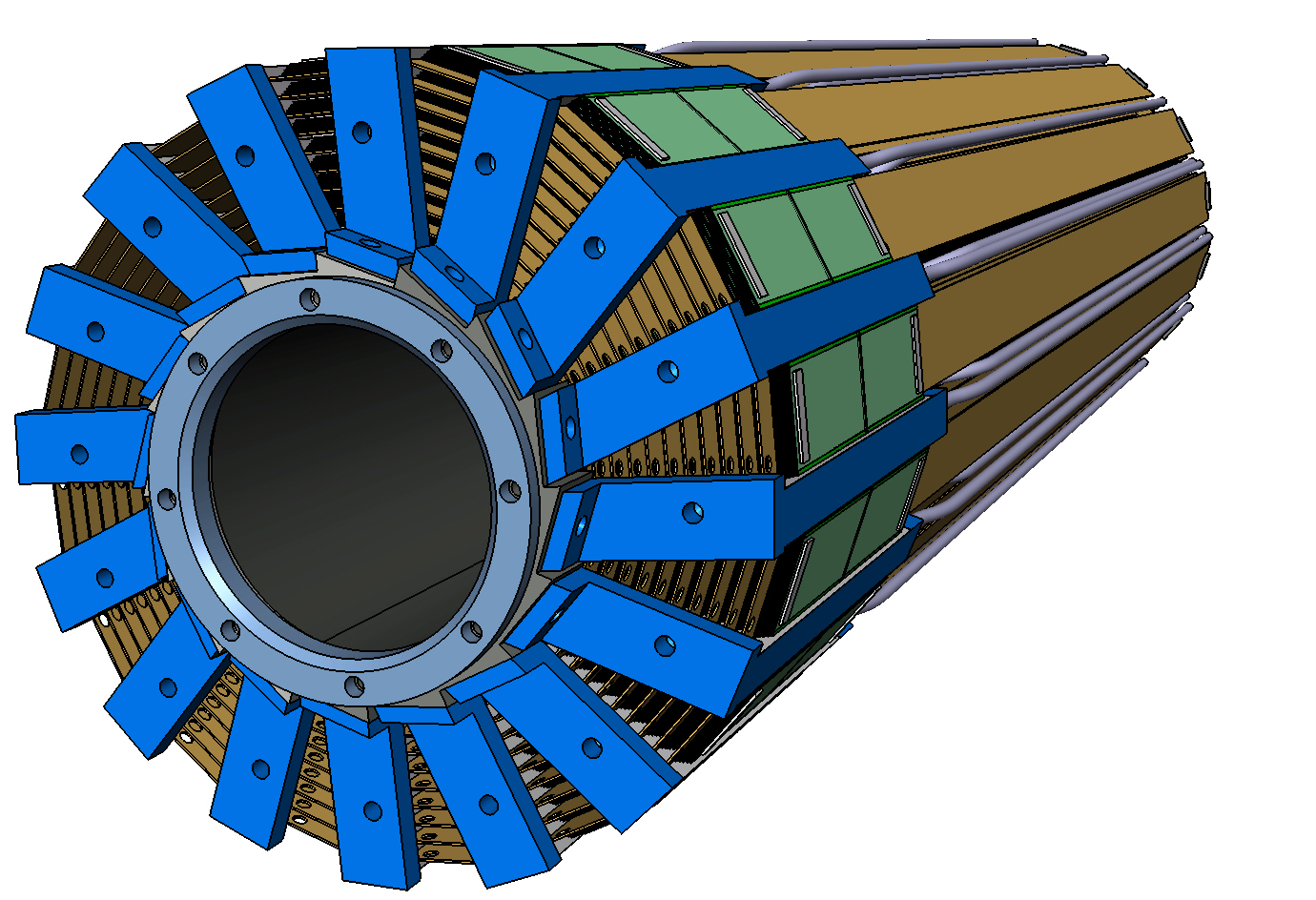}
	\end{minipage}\hfill
	\begin{minipage}{0.50\textwidth}
		\centering
		\includegraphics[width=0.9\textwidth]{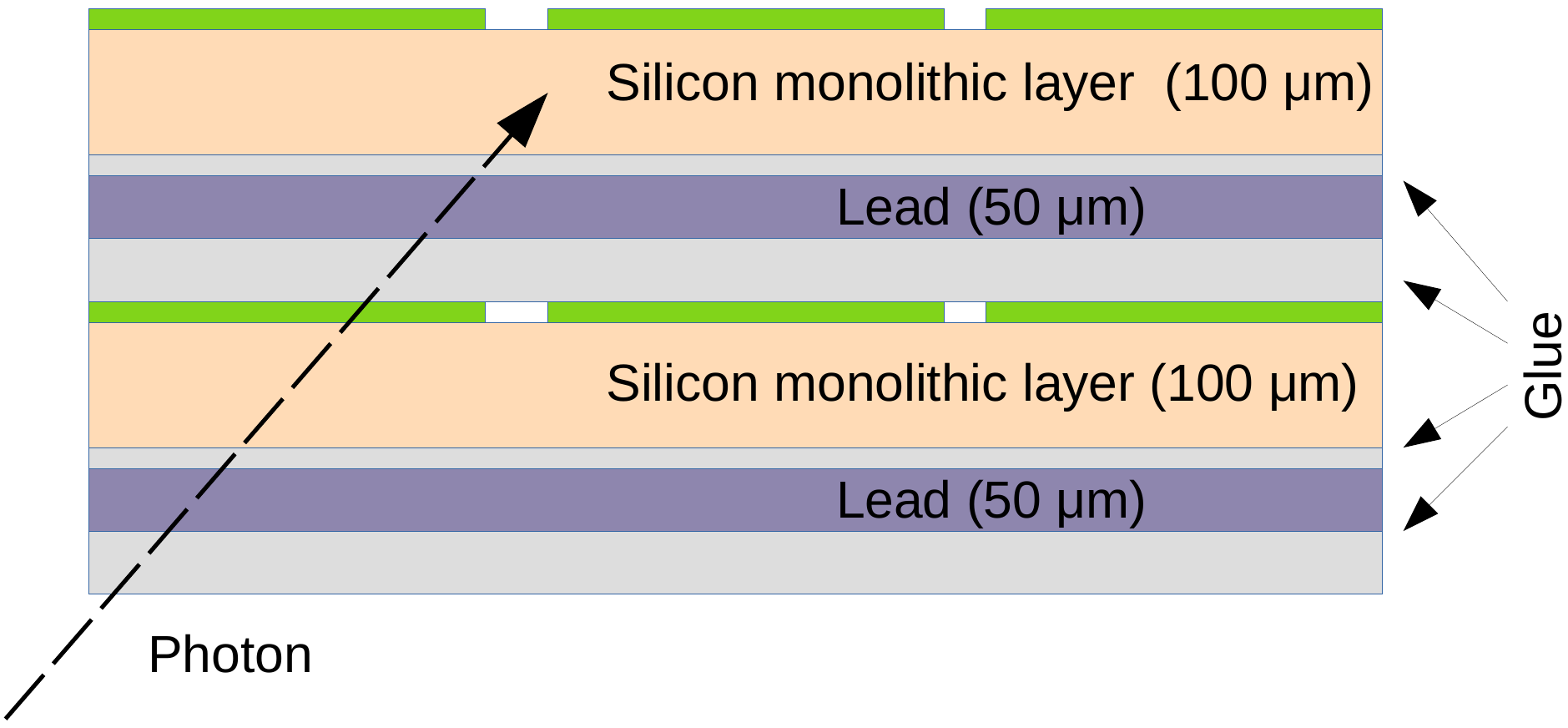}
	\end{minipage}
	\begin{minipage}{0.48\textwidth}
		\centering
		\caption{CAD image of the TT-PET scanner, with the 16 towers and the cooling blocks between them represented in blue. The wedge-shaped towers are formed by ASICs of three sizes, with larger ones at larger radii.}
		\label{fig:barrel}
	\end{minipage}\hfill
	\begin{minipage}{0.48\textwidth}
		\centering
		\caption{Two detection layers, including a monolithic detector, a lead converter and glue. The 60 detection layers of a tower are divided in 12 stacks of 5 layers each (called ``supermodules''). The lead and silicon layers are glued together with \SI{5}{\micro\meter} and \SI{50}{\micro\meter} thick adhesive tape.}
		\label{fig:stack}
	\end{minipage}
\end{figure}

The TT-PET scanner is formed by 16 identical wedges, called towers, containing the detector stack, the mechanical support structures, the cooling and the interconnections (figure \ref{fig:barrel}).
Each detection layer is composed by two \SI{100}{\micro\meter} thick monolithic pixel silicon detectors placed side by side, a \SI{50}{\micro\meter} lead converter and dielectric glue layers, as shown in figure \ref{fig:stack}. Pixels have an area of \SI{500}{\micro\meter} by \SI{500}{\micro\meter}, which corresponds to an input capacitance for the Front-end of about \SI{500}{\femto\farad} including routing.\\
\\
Detectors are grouped every 5 layers in a "super-modules", sharing services and interconnections. The chips in a super-module are all connected to the same flex cable with stacked wirebonds and are daisy-chained to minimize the number of connections needed for the readout.\\
Cooling is provided with a microchannel liquid flow in the blocks between the towers. This solution minimizes the dead area, but it can only dissipate a limited amount of power. Heat transfer simulations by FEA, confirmed by measurements on a mechanical mock-up, were used to calculate the power budget of the detectors, which was set to \SI{200}{\micro\watt} per channel.\\
Three different chip sizes (\SI{25}{\milli\meter} long and 7, 9 or \SI{11}{\milli\meter} wide), are implemented to form wedges. The number of chips was optimized with GEANT4 simulations that allowed the calculation of the scanner sensitivity and efficiency.\\
\section{The TT-PET small-size demonstrator chip}
\label{sec:electronics}
\begin{figure}[htbp]
	\centering
	\includegraphics[width=0.9\textwidth]{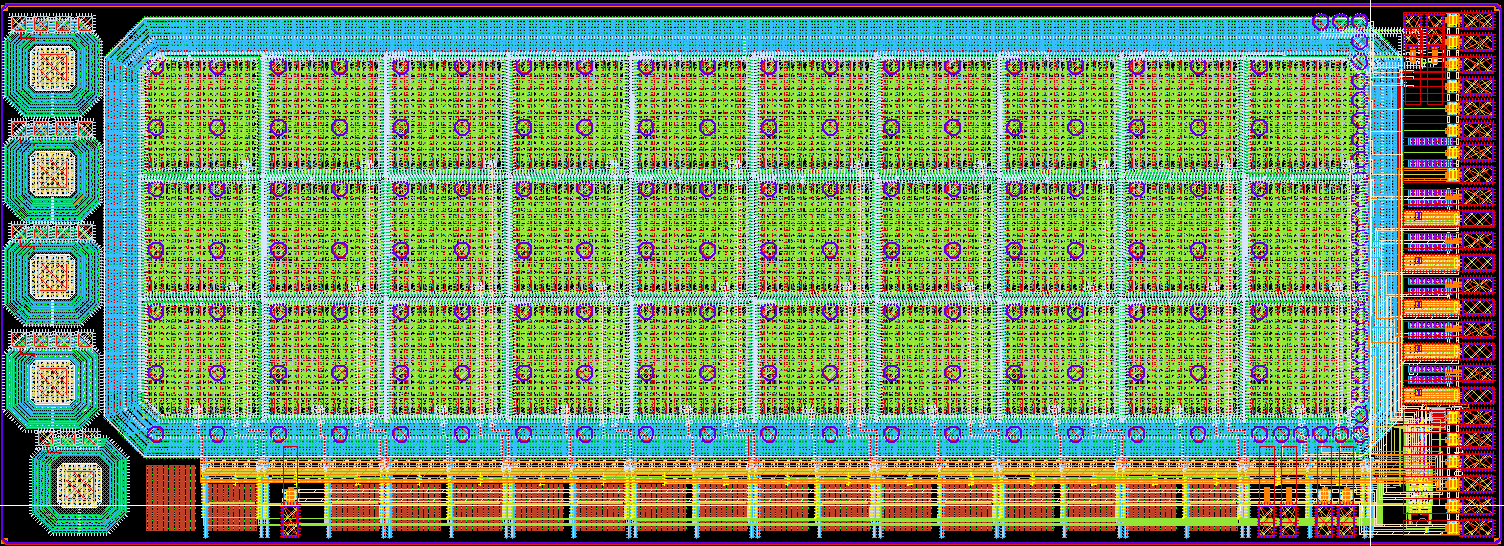}
	\caption{Layout of the TT-PET demonstrator chip, with a 3$\times$10 pixel matrix. On the left, five guard-ring test-structures are visible, that were submitted to independently test the high-voltage insulation of the pixels.}
	\label{fig:chip_figure}
\end{figure}

\begin{figure}[htbp]
	\centering
	\includegraphics[width=0.9\textwidth]{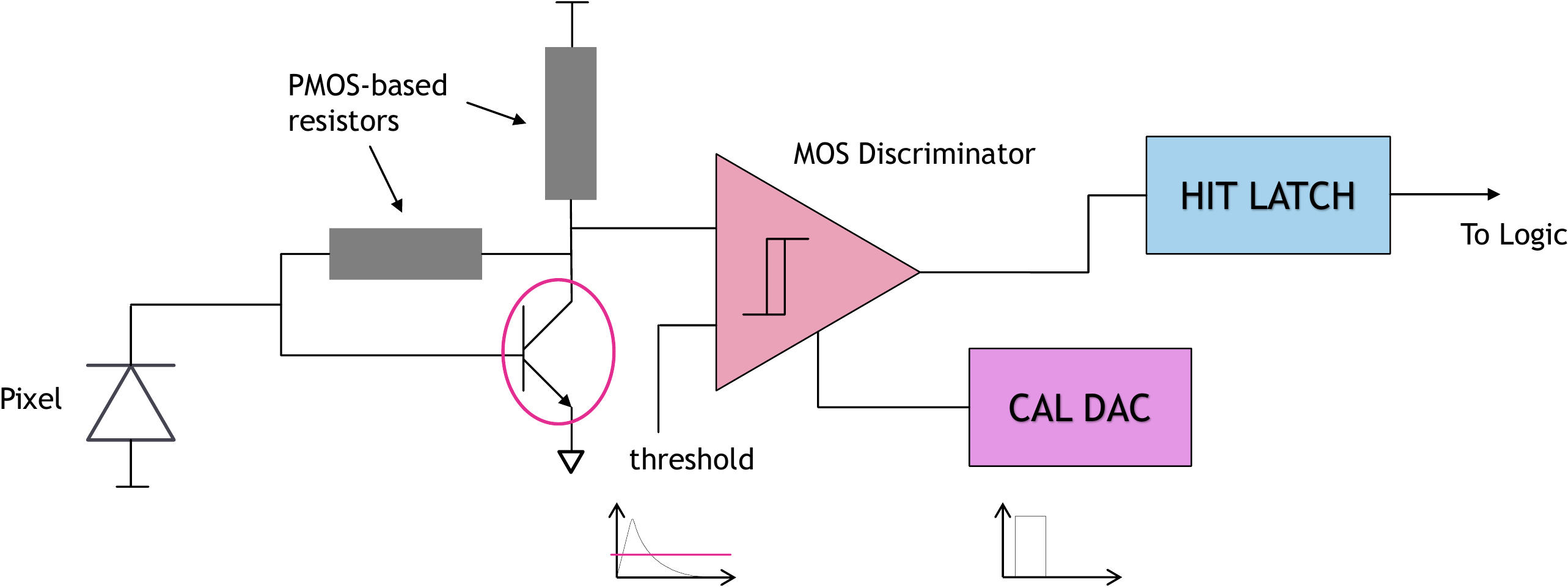}
	\caption{Block diagram of the pixel electronics. The pixel is shown as a diode, connected to the BJT-based preamplifier. Its output is discriminated by an open-loop MOS amplifier, controlled by a local DAC to adjust its threshold. The digitized output is sampled by a latch and sent to the periphery to the TDC.}
	\label{fig:blockdiagram}
\end{figure}

After some small-scale test structures, a 3$\times$10 matrix of fully-featured pixels (shown in figure \ref{fig:chip_figure}) was submitted in a MPW run in Spring 2017. The chip has been fully characterized with radioactive sources and in the SPS beam test facility at CERN.\\
Each of the 500$\times$\SI{500}{\micro\meter} pixels includes a BiCMOS preamplifier, a fast discriminator and an 8-bit calibration DAC for threshold equalization, placed in a column next to the active collection area. In the periphery a single TDC is used to digitize timing information, with all the pixels multiplexed to it. A digital logic block encodes the digitized data along with the hit position and implements a simple I/O protocol for chip readout and configuration. Other blocks include tunable biasing structures for the analog circuits. A block diagram of the pixel electronics can be found in figure \ref{fig:blockdiagram}.
\\

\subsection{Specifications}
\label{sub:specs}
The main specifications for the front-end are shown in table \ref{tab:specs}.
\begin{table}[htb]
	\centering
	\begin{tabular}{|cc|}
		\hline Power supply & \SI{1.8}{\volt} \\ 
		\hline Gain & \SI{90}{\milli\volt\per\femto\coulomb} \\ 
		\hline Equivalent Noise Charge (for a \SI{1}{\pico\farad} input capacitance)& \SI{600}{\elementarycharge\tothe{-}}\\ 
		\hline Power consumption & \SI{135}{\micro\watt} \\ 
		\hline Peaking time & \SI{1.3}{\nano\second} \\ 
		\hline Simulated ToA jitter (for a \SI{1}{\femto\coulomb} signal)& \SI{82}{\pico\second} \\ 
		\hline 
	\end{tabular} 
	\caption{Main specifications of the simulated analog front-end}
	\label{tab:specs}
\end{table}\\
The pixel size is a compromise between input capacitance and power consumption. Having smaller pixels would lead to better spatial resolution of the scanner, but since a PET image has an intrinsic resolution of about \SI{500}{\micro\meter}\cite{petresolution}, the image quality would not improve. A smaller pixel would result in a smaller input capacitance for the amplifier, and thus lower noise, leading to more accurate timing. On the other hand, more channels would be required to cover the same area, so power consumption would increase.\\
Noise is the main contributor to the timing resolution. Given an accurate enough TDC (TDCs with precision of a few \SI{}{\pico\second} can be found in literature\cite{picosecond}), the uncertainty is dominated by the effect of the analog front-end. This includes different factors, such as the pixel-to-pixel threshold variation, the intrinsic electronic noise of the preamplifier and the distribution of charge collection time in the substrate.

\subsection{Front-end design}
\label{sub:frontend}
\begin{figure}[htbp]
	\centering
	\centering
	\includegraphics[width=0.75\textwidth]{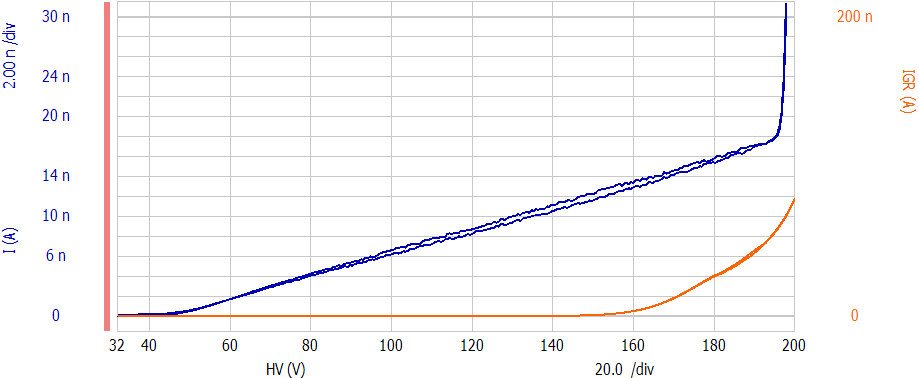}
	\caption{I-V curve of the 3$\times$10 pixel matrix, connecting the backplane to ground and HV through a resistive distribution network. The current going through the diode is in blue (measured when both increasing and decreasing the voltage to show the hysteresis), while the current flowing through the guard ring is in orange.}
	\label{fig:ivcurve}
\end{figure}

\begin{figure}[htbp]
	\centering
	\centering
	\includegraphics[width=0.75\textwidth]{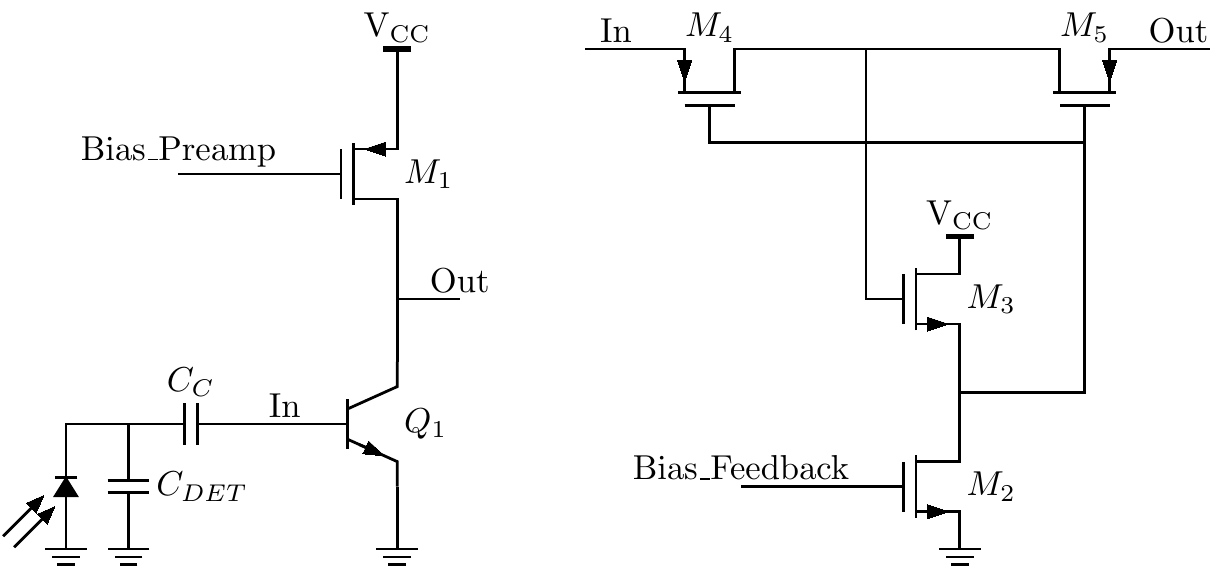}
	\caption{Schematics of the preamplifier. The left block is a common-emitter configuration capacitively coupled to the sensor, while the right one emulates a floating MOS-based feedback resistor which can be tuned from the periphery with a current DAC.}
	\label{fig:preamp}
\end{figure}

\begin{figure}[htbp]
	\centering
	\begin{minipage}{0.50\textwidth}
		\centering
		\includegraphics[width=0.9\textwidth]{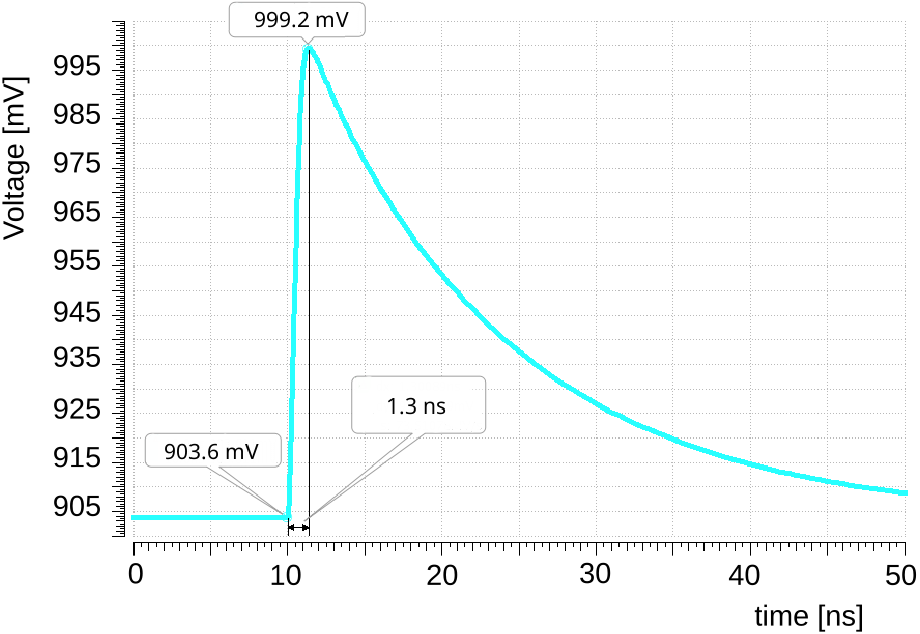}
	\end{minipage}\hfill
	\begin{minipage}{0.50\textwidth}
		\centering
		\includegraphics[width=0.9\textwidth]{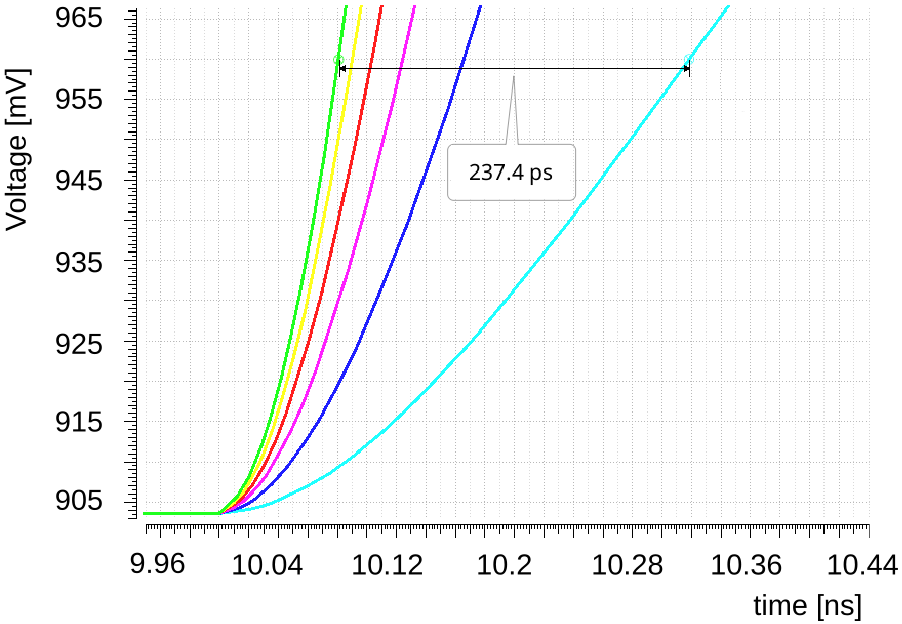}
	\end{minipage}
	\begin{minipage}{0.47\textwidth}
		\centering
		\caption{Typical output of the preamplifier for an input charge of \SI{1}{\femto\coulomb}, extracted from a Cadence Spectre simulation.}
		\label{fig:output}
	\end{minipage}\hfill
	\begin{minipage}{0.47\textwidth}
		\centering
		\caption{Effect of time-walk on the measurement, showing the difference in timing for different input charges, ranging from 1 to \SI{20}{\femto\coulomb}.}
		\label{fig:timewalk}
	\end{minipage}
\end{figure}

\begin{figure}[htbp]
	\centering
	\includegraphics[width=0.45\textwidth]{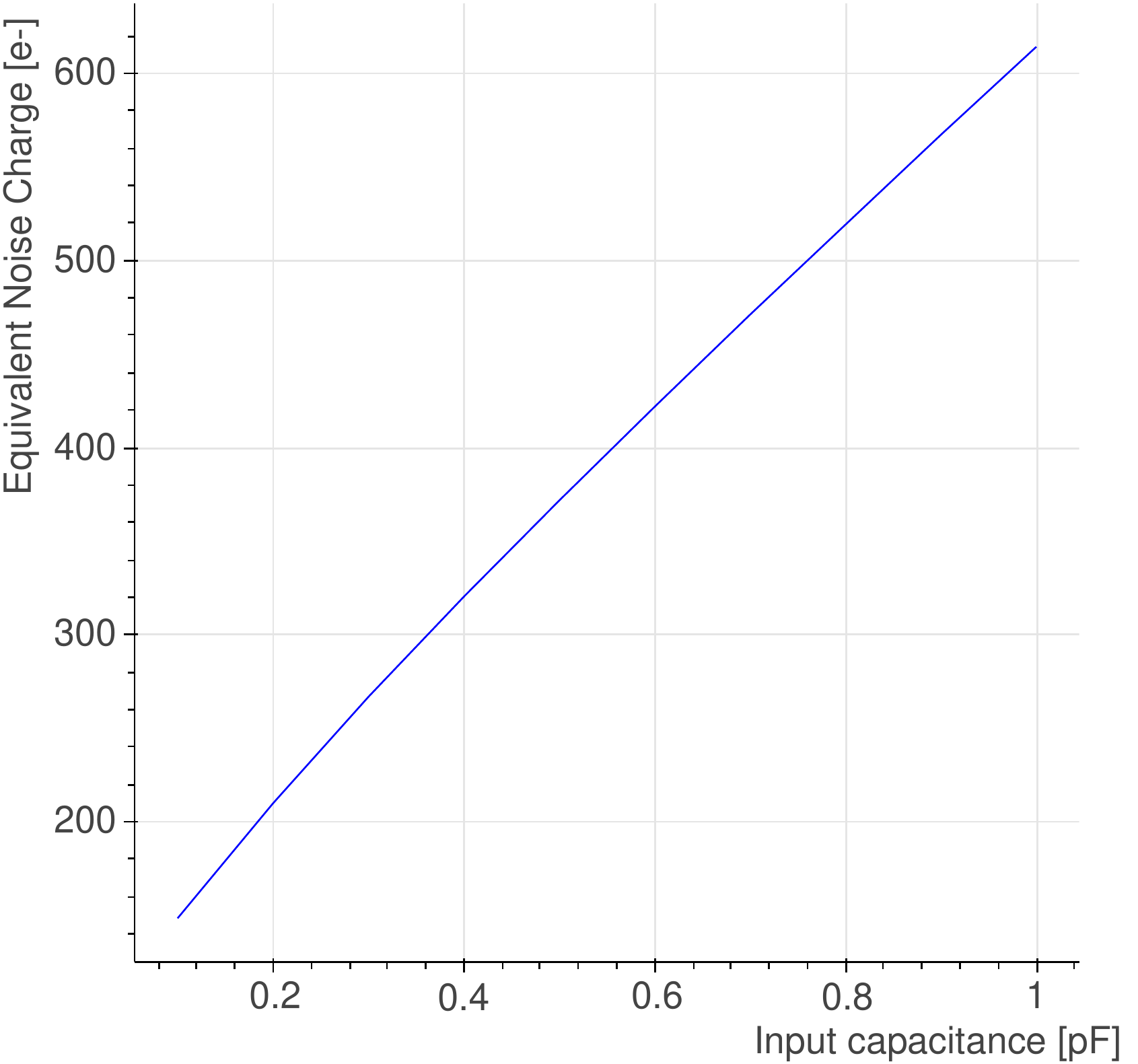}
	\caption{Equivalent Noise Charge of the preamplifier as a function of the input capacitance of the detector. This plot doesn't include the contribution of the discriminator, that filters part of this noise due to its limited bandwidth.}
	\label{fig:enc}
\end{figure}

The front-end features a preamplifier using a Silicon-Germanium Heterojunction Bipolar Transistor (specifically, IHP 130 nm SiGe-HBT technology), which was chosen to minimize the series noise which represents the main contribution to the noise performance.\cite{Paolozzi-thesis}. This front-end was already tested and found to perform well, with the capability of achieving a \SI{100}{\pico\second} jitter for up to \SI{1}{\pico\farad} input capacitance\cite{Paolozzi-100ps}.\footnote{This value is compatible with the target of \SI{30}{\pico\second} for \SI{511}{\kilo\electronvolt} photons. Detailed GEANT4 simulations showed that the average charge deposited by a PET photon would be more than three times larger than the one deposited by a minimum ionizing particle.}\\
The amplifier is connected to the input diode, which is integrated in the electronics substrate, being the chip monolithic. The chip has a \SI{1}{\kilo\ohm} substrate and is thinned to \SI{100}{\micro\meter} in order to optimize the charge collection time and increase the electric field uniformity. Ground reference is provided to the cathode through a back-plane metalization, while the anode is capacitively coupled to the front-end input. Figure \ref{fig:ivcurve} shows the I-V characteristic of the pixel matrix up to a voltage of \SI{200}{\volt}. The leakage current is less than \SI{0.6}{\nano\ampere} per channel, and it is mostly due to the implantation process performed on the backplane. Since the front-end is capacitively coupled to the sensor, the dark current is filtered out and it has a negligible impact on the chip performance.
\\
The preamplifier schematic is shown in figure \ref{fig:preamp}. The BJT is used in a simple common-emitter configuration, with an active PMOS load and a MOSFET feedback, which can be tuned to adjust the equivalent feedback impedance.
\\
The choice of a common-emitter configuration comes from the need to minimize the input and output capacitances to achieve a high gain while keeping the rise time as short as possible. Indeed, the time resolution is directly proportional to the rise time and inversely proportional to the signal-to-noise ratio\cite{Paolozzi-thesis}. This implementation features a 20\%-80\% rise time of about \SI{600}{\pico\second}. Total charge integration time is about \SI{1.3}{\nano\second}, which is compatible with the charge collection time in silicon. A plot of the simulated output of the preamplifier is shown in figure \ref{fig:output}. Due to the much larger peaking time compared to the target time resolution, time walk must be taken into account and compensated when calculating the time of arrival because different input charges can change the time stamp by hundreds of \SI{}{\pico\second}, as shown in figure \ref{fig:timewalk}. This is possible by estimating the charge performing a time-over-threshold measurement and then correcting the time-walk error off-line. Figure \ref{fig:enc} shows the Equivalent Noise Charge referred to the input of the preamplifier for different values of input capacitances.
\\
\\
Each preamplifier is connected to a 3-stage MOS discriminator with a \SI{4}{\milli\volt} hysteresis to compare its output with a fixed threshold. In order to minimize the load capacitance of the amplifier the input stage of the discriminator uses very small NMOS transistors, leading to a significant pixel-to-pixel threshold mismatch (simulations showed a 3\Sigma value of \SI{100}{\milli\volt}). To compensate for this effect, an 8-bit calibration DAC is included in each front-end. It is a binary-weighted, current-steering DAC connected to the first stage of the discriminator that is used to unbalance the current flowing in the two branches and moves the effective threshold of the discriminator. This can compensate for other pixel-to-pixel effects, due for example to the DC output of the preamplifier. The total current produced by the DAC can be tuned to change the calibration range.

\subsection{Readout logic and other blocks}
\label{sub:readout}
Given the low hit rate that we expect in any of the TT-PET chips, all pixels are multiplexed to the same TDC, so that the chip will not be able to detect simultaneous particles. Since this event is very rare\cite{Emanuele-rate}, this approach was chosen to simplify the design and reduce the power consumption of the chip. A single 50-ps binning TDC is placed in the chip periphery and all pixels are connected to it through a balanced ladder of NAND/NOR gates. The TDC measures both time of arrival and time over threshold of the signal, used to compensate for time-walk effects. A separate set of row and column lines are used to extract the pixel address and store it in a readout buffer. Pixel-to-pixel delay, while minimized by the balanced multiplexing network, is still larger than the time resolution, so it requires off-line calibration. The contribution to the time resolution of the digital chain was measured using a testpulse injection circuit and found to be in the order of \SI{1}{\pico\second}.\\
The chip features a simple serial interface for both readout and programming, with data shifted in the pixel configuration memories being connected as a long shift register. Since the chip can only store a single hit at a time, there is a dead time of about \SI{5}{\micro\second} (this value is much )after every hit to allow for the readout of the TDC data. A trigger signal, produced by a fast OR of all the pixels, is also available in output to implement a trigger logic or for debugging purposes.

\section{Results}
\label{sec:results}
\begin{figure}[htbp]
	\centering
	\includegraphics[width=0.7\textwidth]{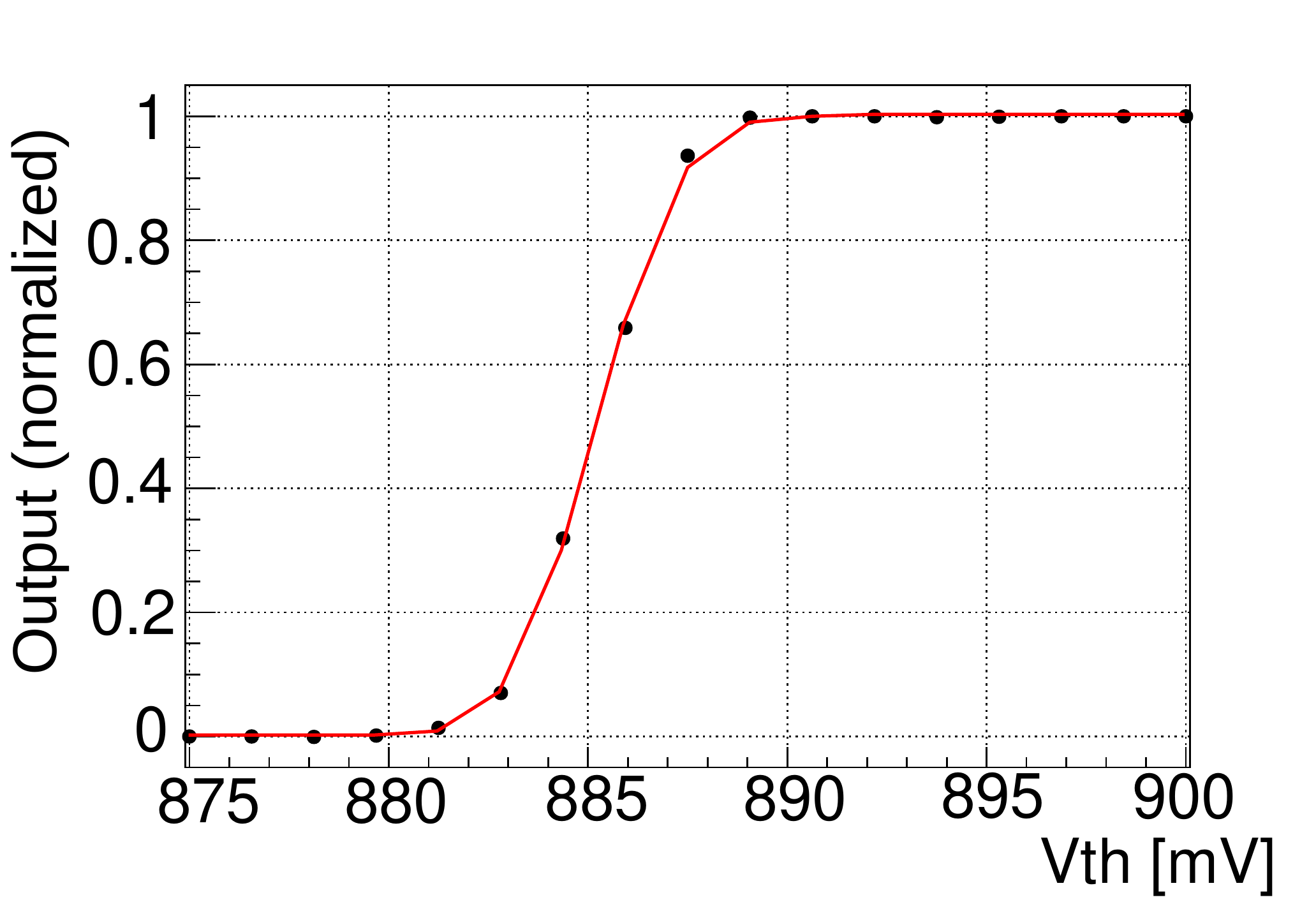}
	\caption{S-curve noise measurement at the output of the discriminator.}
	\label{fig:scurve}
\end{figure}

The demonstrator chip was thoroughly tested with a \ce{^{90}Sr} source at the University of Geneva. For testing purposes the inclusion of a fast trigger signal was very useful as it allowed to characterize and debug the pixel front-end and the TDC separately. The chip is fully working at the nominal power consumption.\\
Noise scans were performed by sweeping the threshold and looking at the real-time output of the fast-OR with an oscilloscope. S-curves (figure \ref{fig:scurve}) were produced and fitted to extract the electronic noise at the output of the preamplifier. The error function fitting the experimental data corresponds to a gaussian curve with a standard deviation of \SI{2.35}{\milli\volt}, corresponding to an input referred noise of less than 400 electrons. It has to be noted that the discriminator has an important impact on this measurement, as it acts as a band-pass filter for the noise. According to Cadence Spectre simulations, the noise standard deviation at the output of the discriminator was reduced by 30\% compared to one at the output of the preamplifier.
\begin{figure}[htbp]
	\centering
	\includegraphics[width=0.7\textwidth]{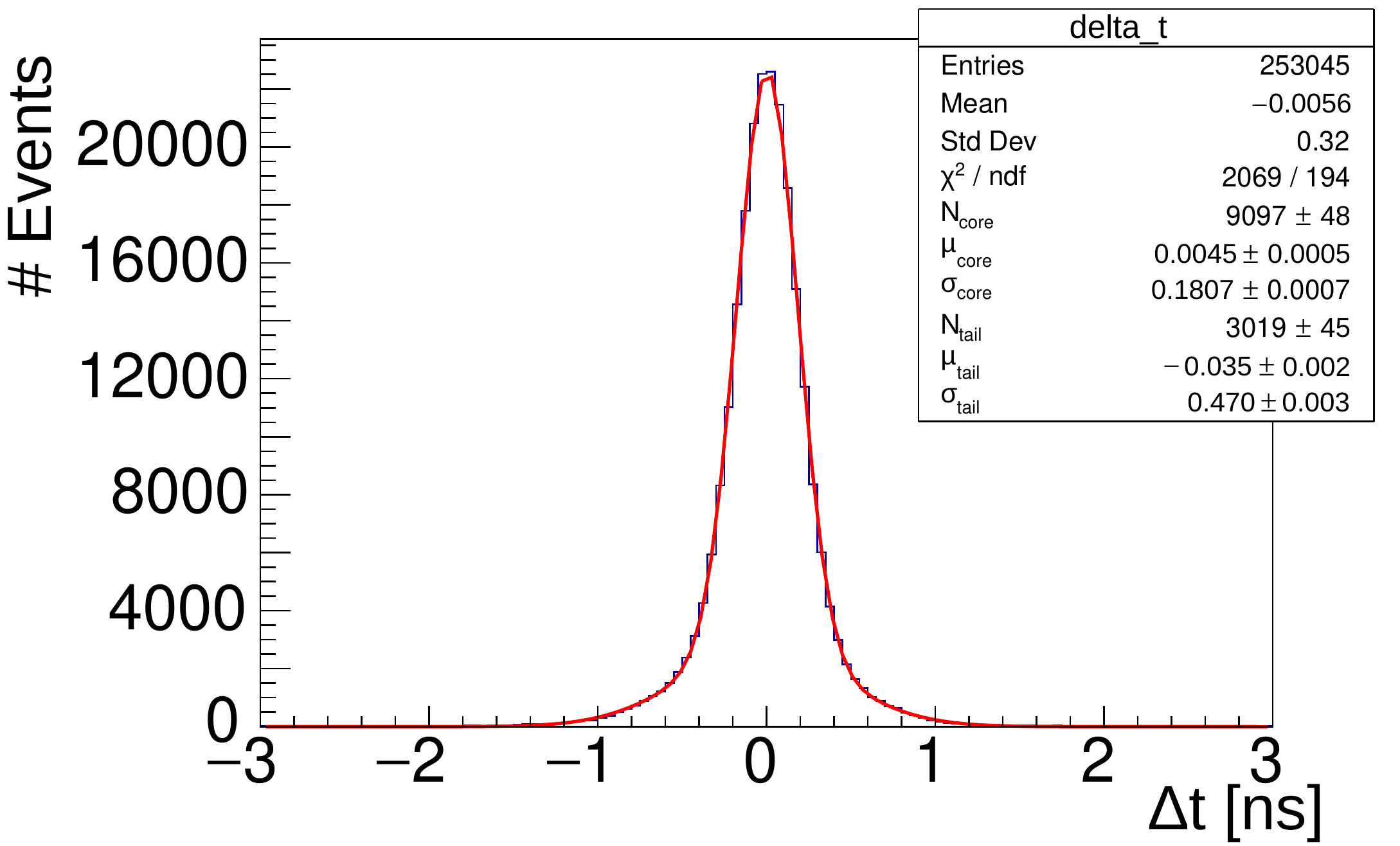}
	\caption{Time-of-flight between two chips obtained with a \ce{^{90}Sr} source. This distribution is fitted with a double Gaussian; the standard deviation $\sigma_{core}$=180$\pm$\SI{0.7}{\pico\second} hints to a time resolution of approximately \SI{130}{\pico\second}, assuming equal performace for the two chips.}
	\label{fig:sr90timing}
\end{figure}\\
Time-of-flight measurements were performed with a \ce{^{90}Sr} source. Two chips were put on top of each other and time differences between them recorded and analyzed. The time-of-flight distribution between the two chips is shown in figure \ref{fig:sr90timing}. The measured time resolution of \SI{130}{\pico\second} for the core of the distribution mis a very promising results, far better than what was previously achieved by monolithic particle detectors.\\
Combined simulations of the sensor and the electronics showed an expected resolution of \SI{92}{\pico\second}. The larger value measured can be attributed to a non-ideal correction for time-walk and to an added input capacitance due to pixel routing, in addition to possible system-level cross-talk from the readout system.\\

\section{Conclusions}
\label{sec:conclusions}
The design of the demonstrator of a monolithic pixel detector for the TT-PET project was presented, together with test results. The chip includes a novel SiGe BiCMOS-based front-end to achieve better than state-of-the-art time resolutions. A time resolution of \SI{130}{\pico\second} was measured with a \ce{^{90}Sr} setup, with a power consumption as little as \SI{135}{\micro\watt} per channel.\\

\acknowledgments

We would like to thank the Electrical Engineering team of the University of Geneva as well as the colleagues of the University of Bern and INFN Tor Vergata for their help with the readout system.
This study was funded by the SNSF SINERGIA grant CRSII2\_160808.


\end{document}